\documentclass[preprint,11pt]{elsarticle}

\usepackage{amsmath}
\usepackage{amsthm}
\usepackage{amsfonts}
\usepackage{hyperref}
\usepackage{cleveref}
\usepackage{booktabs}
\usepackage{placeins}
\usepackage[usenames,dvipsnames]{xcolor}
\newcommand{\ts}[2]{\ensuremath{\mathrm{#1}_{\mathrm{#2}}}}

\newcommand{\kilian}[1]{\textcolor{black}{#1}}

\newcommand{\moritz}[1]{\textcolor{black}{#1}}

\journal{European Journal of Mechanics B/Fluids}

\begin{document}

\begin{frontmatter}



\title{Transient evolution of the global mode in turbulent swirling jets: experiments and modal stability analysis}


\author{Lothar Rukes, Moritz Sieber, Christian Oliver Paschereit, Kilian Oberleithner}

\address{Hermann-F\"ottinger-Institut, TU-Berlin, M\"uller-Breslau-Strasse 8, 10623 Berlin, Germany}

\begin{abstract}
Modal linear stability analysis has proven very successful in the analysis of coherent structures of turbulent flows. Formally, it describes the evolution of a disturbance in the limit of infinite time. In this work we apply modal linear stability analysis to a turbulent swirling jet undergoing a control parameter {\itshape transient}. The flow undergoes a transition from a non-vortex breakdown state to a state with a strong recirculation bubble and the associated global mode. High-speed Particle Image Velocimetry (PIV) measurements are the basis for a local linear stability analysis of the temporarily evolving base flow. This analysis reveals that the onset of the global mode is strongly linked  to the formation of the internal stagnation point. 
Several transition scenarios are discussed and the ability of a frequency selection criterion to predict the wavemaker location, frequency and growth rate of the global mode are evaluated. We find excellent agreement between the linear global mode frequency and the experimental results throughout the entire transient. The corresponding growth rate qualitatively conforms to experimental observations. We find no indication for a nonlinear global mode as proposed by previous studies.
\end{abstract}

\begin{keyword}
linear stability analysis \sep global modes \sep swirling jets \sep transient flows \sep  coherent structures 

\PACS 47.20.-k \sep 47.15.St \sep 47.20.Ft

\end{keyword}

\end{frontmatter}


\section{Introduction}
\label{sec:Intro}
Modal linear stability analysis (LSA) was initially derived for weakly non-parallel laminar base flows, and it describes the evolution of infinitesimal disturbances in the formal limit of infinite time \cite{Huerre.1990}. It has been applied successfully to turbulent mean flows and to flows with varying levels of non-parallelism, \cite{Oberleithner.2011b, Juniper.2011, Gallaire.2003b, Gallaire.2006, Oberleithner.2013}. These studies identified dominant coherent structures with organized stability waves and successfully analyzed these structures. Prime examples are the studies concerning the von K\'{a}rm\'{a}n vortex street \cite{Monkewitz.1988b, Pier.2002, ManticLugo.2015, Thiria.2015} and the spiral global mode of swirling jets undergoing vortex breakdown \cite{Gallaire.2006, Oberleithner.2011b, Oberleithner.2015c}. 
For the latter configuration, it is now well-understood that the flow undergoes a super-critical Hopf bifurcation to a global mode when exceeding a critical swirl intensity. 
Once the flow has bifurcated, the dominant coherent structure is often observed to be a co-rotating single-helical mode with azimuthal wave number $m=1$ \cite{Oberleithner.2012,Liang.2005,Ruith.2003b,Billant.1998}. 
The same structure is frequently observed in swirl flames, where it is termed the precessing vortex core \cite{Syred.2006,Stohr.2012,Oberleithner.2015c}. 

\begin{figure}
\centering
\includegraphics[width= \textwidth]{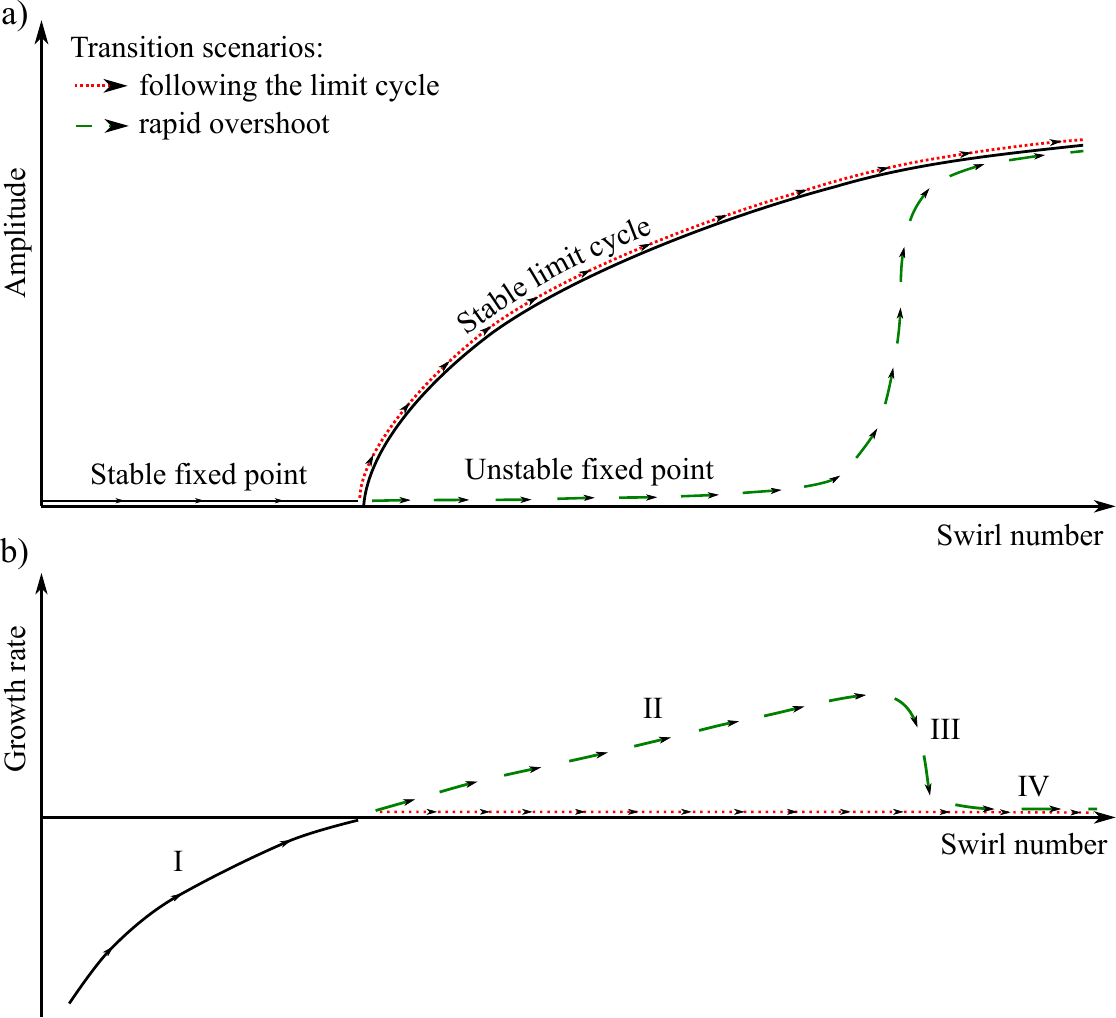}
\caption{(a) Characteristic parabola of a supercritical Hopf bifurcation and different bifurcation scenarios for swirling flows undergoing a control parameter transient. (b) The expected grow rate of the global mode for both transition scenarios. The roman letters indicate characteristic stages of the transient: I -- stable flow, II -- \kilian{rapid base flow transient}, III -- \kilian{growth and} saturation of the mode, IV -- limit cycle oscillation.}
\label{fig:bifurcationDiagramIntro} 
\label{fig:bifDiag} 
\end{figure}

\kilian{Some of these previous studies also focused on flow transients,  considering temporally evolving base flows\cite{Ruith.2003b,Liang.2005,Oberleithner.2015c}}. \Cref{fig:bifurcationDiagramIntro} illustrates the classic parabola of a super-critical Hopf bifurcation. 
By exceeding a critical control parameter, the flow becomes unstable and bifurcates to the oscillating state. 
The aim of transient experiments is to track the amplitude and frequency of the dominant periodic structure during the transient from the unstable (steady) fixed point to the stable (unsteady) limit cycle. 
The beauty of transient flow \kilian{studies} is that they reveal the initial linear mechanism that triggers the instability, as well as the nonlinear saturation process. 

\moritz{We can imagine two distinct bifurcation scenarios depending on the rate at which the control parameter changes (see \Cref{fig:bifurcationDiagramIntro}). 
The path marked by red dots corresponds to a very slow increase of the control parameter (swirl number, where the unstable mode adapts to each change of the boundary conditions and the mode amplitude follows the limit cycle parabola. The global mode growth rate would be negative, when the control parameter is below the critical threshold and approximately zero for supercritical values.
The change in the control parameter, and thus in the flow field, occurs at a time scale that is much slower than that of the instability.
The second scenario involves a rapid increase of the control parameter (swirl number) that does not allow for an intermediate saturation of the instability and the amplitude growth of the global mode lags behind the evolution of the base flow. During this period, the transient base flow should be increasingly unstable to the global mode, which is reflected by a positive growth rate. 
Eventually the global mode saturates at the limit cycle and the growth rates converge to approximately zero. We expect the second scenario for the measurements conducted in this study as will be shown later.
}

\moritz{ Manti\u{c}-Lugo et al. conducted DNS of the cylinder wake and investigated the transient formation of the von K\'{a}rm\'{a}n vortex street \cite{ManticLugo.2015}. They recovered the unstable base flow by artificially constraining the flow to its mid-plane symmetric state. Referring to \cref{fig:bifurcationDiagramIntro}, the unstable fixed point was approached in vertical direction starting from the stable limit cycle. By turning off the constraint and applying a disturbance, the flow "snapped back'' to the limit cycle, undergoing a transient change along the way. The authors conducted a local LSA of the transient base flow and show excellent agreement between the linear global mode frequency and the DNS simulation. They conclude that the mean field modification during the transient causes the shift of the global mode frequency, which confirms the mean field model of Noack et al. \cite{Noack.2003}.
Moreover, they developed a self consistent model of the flow that describes the mean flow deformation and the mode saturation in a quasi static approximation. 
The model employs a steady solution of the Navier-Stokes equations with a forcing term that includes the Reynolds stresses induced by the vortex shedding. 
These stresses are derived from the global linear stability analysis of the quasi steady flow.
The model captures most of the relevant physical mechanisms that are observed in the flow transient from the unstable fixed point to the limit cycle.
Thiria et al. conducted a similar investigation of the transient onset of vortex shedding behind a cylinder \cite{Thiria.2015}. In contrast to Manti\u{c}-Lugo et al., they employed a local linear stability analysis and a subsequent calculation of the frequency selection, which is the same approach pursued in the present paper. Thiria et al. showed that the changing frequencies during the transient, determined from this approach also matches the ones from the DNS. 
}

\kilian{Upon returning our attention to the instabilities in swirling jets,} in the direct numerical simulation of Ruith et al.\cite{Ruith.2003b}, the unstable steady solution was recovered by imposing artificial symmetry constraints. Once these were released, \kilian{a $m = 1$ global mode} arose from a wake in the lee of the breakdown bubble, grew into the bubble, and ultimately set the entire flow into helical oscillation. A LSA of the unstable base flow later associated the dynamics with a nonlinear global mode \cite{Gallaire.2006} that is selected at the upstream transition from convective to absolute instability of the wake. Qadri et al. \cite{Qadri.2013} performed direct numerical simulations of vortex breakdown at different swirl levels. They recover\kilian{ed} the breakdown bubble followed by a wake and identif\kilian{ied} the wavemaker of a nonlinear global mode in the wake region. However, they also report that regions around the bubble are more influential in determining the frequency and growth rate of a linear global mode. 

Liang \& Maxworthy \cite{Liang.2005} conducted an  experimental study of a swirling jet transient. In this study the swirl number was increased transiently, which translates in \cref{fig:bifurcationDiagramIntro} to a horizontal approach of the unstable fixed point. The authors observed that before the onset of vortex breakdown a mode with azimuthal wave number $m = 2$ was dominant, whereas the spiral global mode with $m = 1$ was dominant after vortex breakdown. They suggested that the global mode after vortex breakdown was related to the existence of a region of absolute instability and discussed the possibility that a so-called wavemaker is located upstream of the recirculation bubble, enforcing its global frequency on the entire flow. Although, the existence of a self-excited mode with azimuthal wave number $m=2$ is plausible based on local LSA reasoning \cite{Gallaire.2003b}, there are numerous examples where the $m=1$ mode is the only dominating \cite{Oberleithner.2012,Oberleithner.2014,Stohr.2012,Terhaar.2014b}.

In this study, we investigate a swirling jet transient following the experimental approach of Liang \& Maxworthy. We adopt the theoretical approach of Thiria et al. \cite{Thiria.2015} as well as Manti\u{c}-Lugo et al. \cite{ManticLugo.2015} and apply LSA to the swirling jet base flow that we obtain from high-speed flow measurements. 
The flow considered here is highly turbulent, strongly nonparallel, and features drastic changes during the transient. 
The results from the LSA will be compared to the experimental findings. 
The success of the LSA for this difficult flow configuration will not only support the suggestions made by Liang \& Maxworthy regarding the origin of the global mode, but it will also give credibility to the general analytic approach.
With this work, we intend to answer the following questions:
\begin{itemize}
\item
\textit{Can we apply LSA to a transient swirling flow?}
\item
\textit{\kilian{Where} does the global $m = 1$ instability arise?}
\item
\moritz{\textit{\kilian{What transition scenario applies to the transient experiment?}}}
\end{itemize} 
\moritz{
The manuscript is organized in the following way: \Cref{sec:expSetup} introduces the swirling jet facility, together with the relevant characteristic numbers and our experimental setup. \Cref{sec:dataAnalysis} presents the experimental and numerical methods and \cref{sec:results} describes the obtained results. \Cref{discuss} discusses the transient evolution of the flow and the results of the spatio-temporal analysis and \cref{concl} gives a conclusive summary.
}

\section{Experimental setup}
\label{sec:expSetup}

\subsection{The swirling jet facility}
\begin{figure}
\centering
\includegraphics[width= \textwidth]{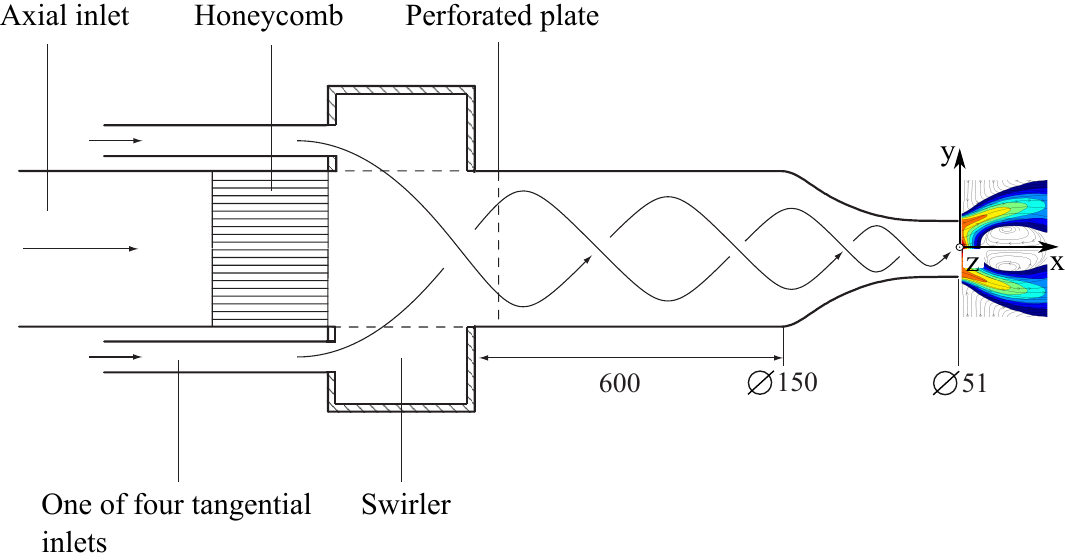}
\caption{The experimental facility. All dimension are in millimeters and not drawn to scale.}
\label{fig:expFacility} 
\end{figure}

The swirling jet facility was inspired by the experimental apparatus of Chigier \& Chervinsky \cite{Chigier.1965}. The primary axial air stream was passed through a deep honeycomb before it entered  the swirler through a single concentric inlet. Swirl could be imparted on the primary air stream through a secondary air stream that entered the swirler through four tangential slots, which had a length of 80 mm each. Both air streams were supplied by separate frequency controlled blowers. The volume flux of each air stream was monitored and controlled by calibrated orifices. Downstream of the swirler, the swirled air passed through a 600 mm long tube, followed by a contracting nozzle and exited into the unconfined surrounding. The nozzle had an outlet diameter, D, of 51 mm and a contraction ratio of 9. The swirl velocity of the unconfined jet was set by the ratio of the primary to the secondary volumetric flow rate. A non-swirling round jet was generated when the volumetric flow rate of the secondary air stream was set to zero. The largest swirl levels were attained when  the primary blower was not operated and the secondary blower supplied its largest volumetric flow rate. Further details on this experimental facility are provided in the studies of Oberleithner et al.\cite{Oberleithner.2011b,Oberleithner.2014}. \Cref{fig:expFacility} also introduces a Cartesian coordinate system with its origin placed on the jet axis in the nozzle exit plane. The $x$-axis is oriented in the flow direction, the $y$-axis in the cross-flow direction and the $z$-axis in the out-of-plane direction. Additionally, a cylindrical coordinate system is introduced with the same origin. The $r$-axis is aligned with the $y$-axis of the Cartesian system at zero degrees of revolution, $\theta$ is counted mathematically positive and $x$ points in the streamwise direction. The velocity vector $\textbf{v}$ has the components $v_{x}$, $v_{y}$ and $v_{z}$ in the Cartesian system, and $v_{x}$, $v_{r}$, $v_{\theta}$ in the polar coordinate system.

\subsection{Characteristic numbers}
The flow is characterized by the swirl number \cite{Chigier.1967}, 
\begin{equation}
	\mathrm{S} = \frac{\dot{G_{\theta}}}{\mathrm{D}/2  \dot{G_{x}}} = \frac{2\pi  \int\limits_{0}^{\infty}\rho  \overline{v}_{x} \overline{v}_{\theta}  r^{2} \mathrm{d}r}{\mathrm{D}  \pi \int\limits_{0}^{\infty}\rho \left(\overline{v_{x}^{2}}- \frac{\overline{v_{\theta}^{2}}}{2}\right)  r \mathrm{d}r},
	\end{equation}
which is measured at the most upstream available position, $x/\mathrm{D} = 0.12$. It quantifies the amount of swirl in the flow and is defined as the ratio between the axial flux of azimuthal momentum $\dot{G}_{\theta}$ and the axial flux of axial momentum $\dot{G}_{x}$. The swirl number ranged between 0.85 at the beginning of the transient and 1.3 at the end of the transient. The total time of the transient was 3 s. The Reynolds number
\begin{equation}
\mathrm{Re} = \frac{\ts{v}{bulk} \mathrm{D}}{\nu},
\end{equation}
with $\ts{v}{bulk}$ indicating the bulk flow velocity, D the nozzle diameter and $\nu$ the kinematic viscosity of air. With a bulk velocity of 6.3 m/s, the Reynolds number amounted to 21000 and was constant during the transient.

\subsection{Data acquisition and experimental procedure}
The properties of the measurement system are discussed in the following. A high-speed PIV system was used for the investigation of the swirl transient. The high-speed stereo system featured two CMOS cameras with a resolution of 1000x1000 pixel and a laser that was operated at 500 Hz. Data was acquired for a total time of 3 s, resulting in 1500 snapshots. 

The double images were processed using the commercial software VidPIV using standard digital PIV processing \cite{Willert.1991}. The data analysis employed iterative multigrid interrogation with image deformation \cite{Scarano.2002}. The final size of the interrogation window was 32x32 pixel with an overlap of 50\%. Errors in the laser sheet alignment were minimized by the use of corrected mapping functions. The initial datum calibration marks were back-projected onto the measurement plane by an optimized Tsai camera model \cite{Soloff.1997}.

The swirling jet transient was generated by changing the set point of the flow rate at the primary and secondary inlet. 
During the transient, the total flow rate was kept constant and only the relative contributions of the primary and secondary air stream were altered.
The two blowers had low response times, therefore any signal obtained from them (such as the rotation rate) was unsuitable to time the beginning of data acquisition. Instead, the mass flow rate measurements at the two orifices were used to determine the exact time when to start the acquisition of the transient event. Prior to the PIV measurements, the ratio of the axial and azimuthal flow rates was determined, at which vortex breakdown occurs. When this critical ratio was reached a trigger impulse was sent from the volume flow control system to the PIV acquisition system.

\section{Methods}
\label{sec:dataAnalysis}
\subsection{Decomposition of the transient flow data}

Turbulent flows that feature a dominant coherent structure are typically decomposed into three parts: the mean, the coherent, and the turbulent part \cite{Reynolds.1972}. Accordingly, for flows undergoing transient dynamics, the modification of the mean flow would be lumped into the turbulent part. It is, therefore, more appropriate to decompose the transient data into four parts: the mean, the shift, the coherent and the turbulent part, reading 
\begin{equation}\label{eqn:triDeco}
\textbf{v}(\textbf{x},t) = \overbrace{\underbrace{\overline{\textbf{v}}(\textbf{x})}_{\mathrm{ mean}}+\underbrace{\widetilde{\overline{\textbf{v}}}(\textbf{x},t)}_{\mathrm{shift}}}^{\text{transient mean flow }\overline{\textbf{v}}^{\dagger}} + \underbrace{\widetilde{\textbf{v}}(\textbf{x},t)}_{\mathrm{coherent}} + \underbrace{\textbf{v}'(\textbf{x},t)}_{\mathrm{turbulent}}.
\end{equation}
We utilize proper orthogonal decomposition (POD) to separate the transient and coherent part from the turbulent fluctuations. 
The transient part will be represented by two POD shift modes that capture the long-term drift of the mean flow during the transient \cite{Noack.2003}. 
\moritz{
The timescales represented by the shift mode are considerably smaller than the period of the coherent structure. On the limit cycle, the coherent mode has a peak frequency of approximately 50Hz whereas the frequency content of the shift mode lies below 10Hz. Therefore, the decomposition of the flow in terms of timescales is a low frequency part describing the transient mean flow, an intermediate frequency part represented by the coherent structure and a high frequency part that consists of stochastic turbulent fluctuations.
}

The coherent (periodic) part is captured by two temporally coupled POD modes of equal energy \cite{Oberleithner.2011b}. 
The methodology is outlined in detail in related previous publications dealing with natural and forced swirling jets \cite{Oberleithner.2011b,Oberleithner.2011,Rukes.2015}. In what follows, we briefly outline the POD-based approach.

POD allows for an efficient characterization of the flow dynamics in a orthogonal subspace that is optimal in terms of the captured kinetic energy \cite{Berkooz.1993}. The starting point of the POD is a decomposition of the velocity vector \textbf{v} into a mean $\overline{\textbf{v}}$ and a fluctuating part $\widehat{\textbf{v}}$,
\begin{equation}
\textbf{v}(\textbf{x},t) = \overline{\textbf{v}}(\textbf{x}) + \widehat{\textbf{v}}(\textbf{x},t).
\end{equation}
The fluctuating part is decomposed in a mode expansion, 
\begin{equation}
\label{eqn:PODstandard}
\widehat{\textbf{v}}(\textbf{x},t) = \sum_{i=1}^{N_i}{a_{i}(t)\boldsymbol{\Psi}_{i}(\textbf{x})}+\textbf{v}_{\mathrm{res}},
\end{equation}
and the residual $\textbf{v}_{\mathrm{res}}$ is minimized for any subspace $N_i<N$ utilizing a least-squares approach, where N is the number of snapshots. The $a_{i}(t)$ is the temporal coefficient of the $i$th POD mode $\boldsymbol{\Psi}_{i}$. 

\begin{figure}
\centering
\includegraphics{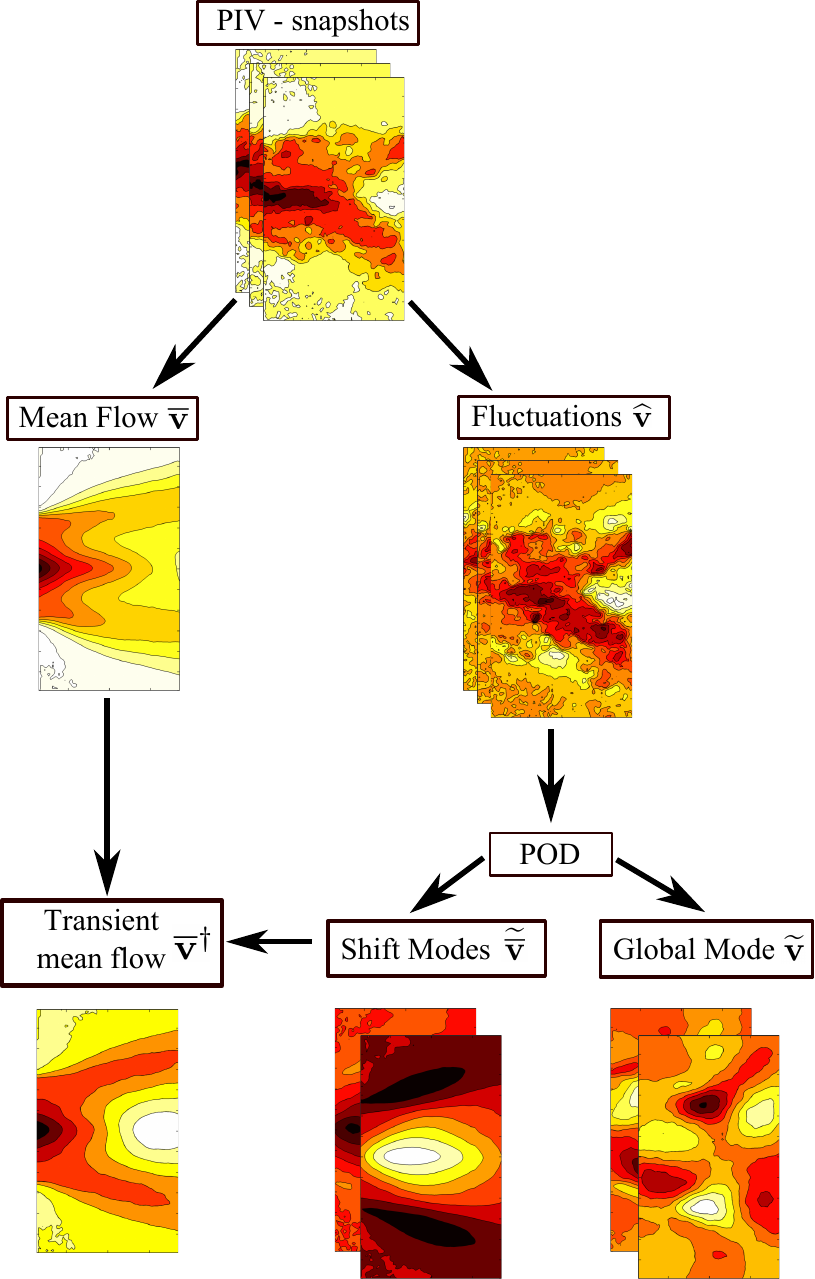}
\caption{Outline of the POD analysis. The mean flow, shift modes and global modes are identified from PIV snapshots. The combination of mean flow and shift modes provides the input for the LSA.}
\label{fig:PODoutline} 
\end{figure}

The velocity decomposition is outlined in \Cref{fig:PODoutline}. Following the flow chart from top to bottom, the PIV snapshots yield a time-mean field and instantaneous fields of velocity fluctuations. The latter are decomposed in POD modes and shift and coherent parts are identified by their respective symmetries, energies, and temporal coefficients. The shift modes are axisymmetric, highly energetic, and feature non-periodic time coefficients that follow the flow transient. The coherent modes are anti-symmetric (m=1), of equal energy, and their temporal coefficients show clear periodicity. 
All remaining modes are considered as turbulence and lumped into the stochastic part. 
The shift modes are added to the mean field to obtain the transient mean flow as indicated in \cref{eqn:triDeco}.
\subsection{Determining the transient global mode amplitude and frequency}
\label{AmplAndFreq}

The POD outlined in the last section is a convenient tool to extract structures that change globally over the course of the transient. It is however not well suited to extract instantaneous information about the global mode. Short time fast Fourier transform (FFT) is ideally suited for this task. The results of the short time FFT complement the POD analysis of the experimental data and provide the benchmark for the LSA.\\
Only velocity fluctuations related to the global mode are considered in the short time FFT. The global mode has an azimuthal wave number of $m = 1$ and related velocity fluctuations need to have a sign change across the jet center line. Only this pre-filtered anti-symmetric part of the velocity fluctuations is considered.

The short time FFT of the pre-filtered velocity signal is computed at every single spatial point and for all three velocity components.
The FFT is applied to a segment of 128 ($2^7$) samples which corresponds to approximately 13 periods of the global mode oscillation.
The magnitude of the FFT spectrum is integrated from $25$ to $100$ Hz and summed-up over all velocities, providing a single amplitude at each spatial point.
This gives an instantaneous (at the center of the time segment) measure for the spatial amplitude distribution of the global mode.
The frequency band was chosen to accommodate the change in frequency of the global mode during the transient, where the mean frequency of the global mode in this transient is approximately 50 Hz.

In addition to the amplitude, the instantaneous frequency of the global mode is determined from the short-time FFT. The histogram of frequencies in the range from 25-100 Hz is considered at each spatial point of the vector field at each time step. The frequency with the largest histogram count is then selected at each point \moritz{(the frequencies over all spatial points are counted)}. The frequency of the global mode at each time step is the frequency that was most often picked from the histogram count at that time step. \moritz{The reasoning behind this procedure is that a global mode would synchronize the oscillations in the entire flow. Therefore, we picked the frequency that dominates the largest part of the flow to be the global modes frequency.}

\subsection{Spatio-temporal stability analysis of the transient mean flow}
\label{sec:spatio-tempo}

The governing equations of local LSA are derived by substituting the decomposition $\textbf{v} = \overline{\textbf{v}}^{\dagger} + \widetilde{\textbf{v}}$ into the Navier-Stokes and continuity equations. Linearizing around the transient mean state results in the stability equations 
\begin{align}
\frac{\partial \widetilde{\textbf{v}}}{\partial t} + (\widetilde{\textbf{v}}\cdot \nabla) \overline{\textbf{v}}^{\dagger} +   
(\overline{\textbf{v}}^{\dagger}\cdot\nabla) \widetilde{\textbf{v}} &= -\nabla\widetilde{p} + \frac{1}{\mathrm{Re}}\Delta\widetilde{\textbf{v}} \label{eqn:nsLin}\\
\nabla\cdot\widetilde{\textbf{v}} &= 0\label{eqn:contiLin}.
\end{align}

\moritz{ We assume that the transient mean state $\overline{\textbf{v}}^{\dagger}$ is a quasi steady solution of the forced Navier Stokes equations, as pursued by Manti\u{c}-Lugo et al. \cite{ManticLugo.2015}. Therefore, the temporal changes of the mean flow are contributed to changes of the forcing by the Reynolds stresses. Hence, the timescale at which the mean flow changes is directly coupled to the timescale at which the coherent structure $\widetilde{\textbf{v}}$  is amplified.}

The LSA targets the coherent velocity part, by assuming a decomposition into normal modes
\begin{equation}
[\widetilde{\textbf{v}},\widetilde{p}] = [iF(r),G(r),H(r),P(r)]\mathrm{e}^{i(\alpha x + m\theta -\omega t)}.
\end{equation}
The radial distribution of the disturbance is given by the amplitude functions $F, G, H$ and $P$. $\alpha$ denotes the axial wave number, $m$ the azimuthal wave number and $\omega$ the frequency. \Cref{eqn:nsLin} and \cref{eqn:contiLin} together with the disturbance ansatz and appropriate boundary conditions yield an eigenvalue problem that is discretized with the Chebyshev pseudospectral collocation technique \cite{Khorrami.1989}. The resulting matrix eigenvalue problem is solved in MATLAB for $\alpha$ real and complex, yielding a temporal, respectively spatio-temporal analysis. The spatio-temporal analysis implies searching for saddle points in the mapping from complex $\alpha$ to complex $\omega$ \cite{Huerre.1990}. The absolute frequency $\omega_{0}$ is given at the location $\alpha_{0}$ of the saddle point in the complex plane. The absolute growth rate and absolute frequency are determined by $\Im(\omega_{0})$ and $\Re(\omega_{0})$, respectively. The flow is absolutely unstable, if $\Im(\omega_{0})>0$ and convectively unstable, if $\Im(\omega_{0})<0$. The distinction between convective and absolute instability at every flow slice is an important information, because a finite region of absolute instability is necessary for the flow to sustain a global mode \cite{Chomaz.1988}.

When the presence of a region of absolute instability is established, the properties of the global mode can be derived from a frequency selection criterion. The criterion of \cite{Chomaz.1991} was found to predict the linear global mode frequency in self-excited flows \cite{Pier.2002,Khor.2008,Leontini.2010}. This criterion identifies the global mode growth rate and frequency from the condition
\begin{equation}
\omega_{g} = \omega_{0}(X_{s}), \qquad \text{where } \qquad  \frac{\mathrm{d}\omega_{0}(X_{s})}{\mathrm{d} X} = 0.
\end{equation}  
The frequency selection criterion identifies a saddle point at $X_{s}$ from the analytic continuation of the $\omega_{0}$ curve into the complex $X$-plane. The location $X_{s}$ refers to the wavemaker that imposes its frequency on the entire flow. If $\Im(\omega_{g})>0$, the flow is globally unstable and oscillates at $\Re(\omega_{g})$. If $\Im(\omega_{g})<0$, the flow is globally stable.

\begin{figure}
\centering
\includegraphics{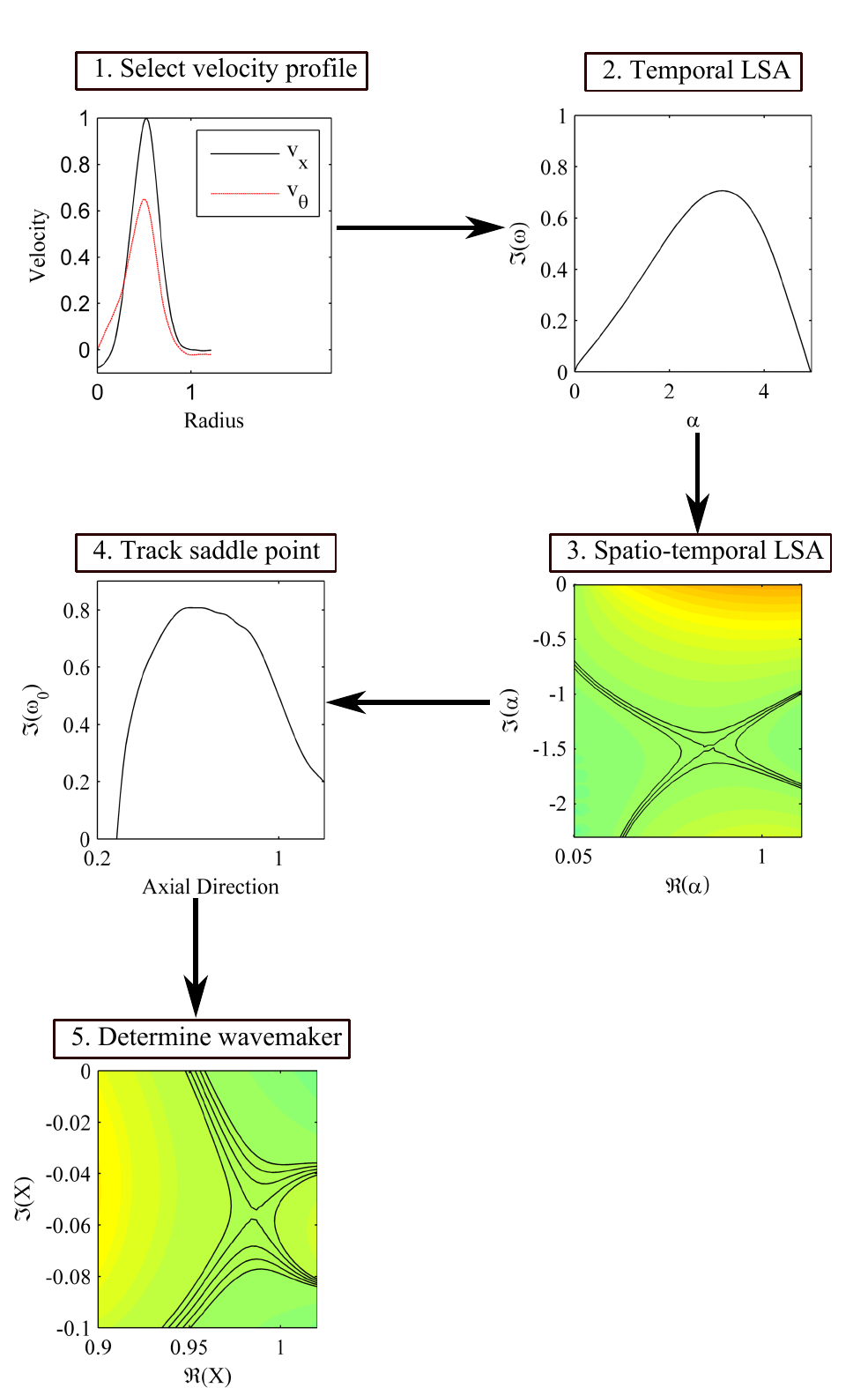}
\caption{Outline of the stability analysis procedure.}
\label{fig:LSAoutline} 
\end{figure}

\Cref{fig:LSAoutline} gives the outline of the entire procedure for obtaining stability analysis results. The path begins in the top left corner of \cref{fig:LSAoutline} and follows the arrows. The following steps are taken:
\begin{itemize}
\item[1.]
Select one velocity profile at one time instance in the transient. A profile that exhibits a substantial velocity deficit, but leaves enough space between the nozzle and the upstream stagnation point is a convenient choice.
\item[2.]
Perform a temporal stability analysis; The axial wave number $\alpha$ is given real, the temporal eigenvalue $\omega$ is computed as a complex number. The choice of velocity profile facilitates the identification of the stability mode that grows on the inner shear layer.
\item[3.]
Perform a spatio-temporal analysis; The axial wave number is complex. $\omega_{0}$ is identified at saddle points over the complex $\alpha$ plane and the validity of the saddle point can be ensured by the cusp map \cite{Kupfer.1987}.
\item[4.]
Perform a spatio-temporal analysis at every other flow slice of the same time instance. \cite{Rees.2009} proposed a convenient procedure to track saddle points along $x$ in an automated way.
\item[5.]
The frequency selection criterion of \cite{Chomaz.1991} is applied. Pad\'{e} polynomials are suitable for the analytic continuation of the absolute growth rate curve into the complex $X$-plane \cite{Juniper.2011}.    
\end{itemize}
\newpage
\section{Results}
\label{sec:results}
\subsection{The transient evolution of the flow}
\begin{figure}
\centering
\includegraphics{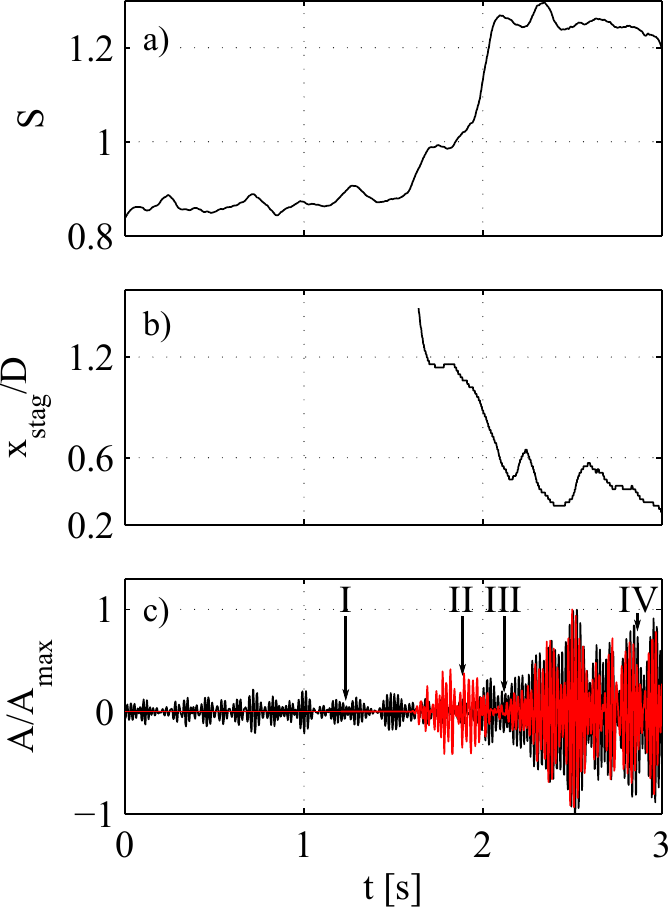}
\caption{Transient evolution of the swirl number, a), and the location of the upstream stagnation point, b). The POD amplitude of the global mode (black) and the radial velocity fluctuation at the stagnation point (red) versus time are presented in c). Roman numerals indicate representative time instances.}
\label{fig:swirlNumber} 
\end{figure}

\Cref{fig:swirlNumber} provides an overview of the flow transient. \Cref{fig:swirlNumber} a) shows the swirl number derived at the streamwise location $x/D=0.12$, which is considered as the control parameter of the transient experiment. \Cref{fig:swirlNumber} b) shows the streamwise location of the upstream stagnation point of the breakdown bubble. It serves as the indicator for the existence and location of vortex breakdown.
Considering these two measures, the transient can be separated into three phases. From zero to 1.5 s, the swirl number remains approximately constant and below the critical threshold for vortex breakdown. From 1.5 to 2.2 s, the swirl number increases rapidly. At the beginning of this period, the first occurrence of vortex breakdown is detected in the measurement domain. After 2.2 seconds, the upstream stagnation point is located close to the nozzle. For times larger than 2.2 seconds, the swirl number levels of. \Cref{fig:swirlNumber} clearly shows that the upstream shift of the stagnation point is directly related to the increase of the swirl number. \Cref{fig:swirlNumber} c) marks four time instances by roman numerals that are representative for different phases of the transient.  
The onset of the global mode oscillations is depicted in \cref{fig:swirlNumber} c) using two independent measures. The black line corresponds to the time coefficient of the POD mode that represents the global mode. The red line corresponds to the radial velocity fluctuations measured at the stagnation point $x_{\mathrm{stag}}$. Note that the amplitude of the POD mode increases somewhat later than determined from the local velocity measure. This is due to the fact that the global mode in its initial stage of growth has a different shape and frequency than at its limit cycle, which is represented by the spatial POD mode. The difference between the linear stability modes of the base flow and the limit cycle oscillations is attributed to the mean flow corrections involved during the saturation process \cite{Noack.2003}. The local velocity measure provides a better estimate of the global mode onset, but it suffers from jitter in the flow transient data. It detects the onset of the global mode at approximately $1.8~$ s, but shows a significant dent in the amplitude later on that is not seen by the POD. Hence, local and global measures should both be consolidated to get an impression of the global mode transient.

\begin{figure}
\centering
\includegraphics{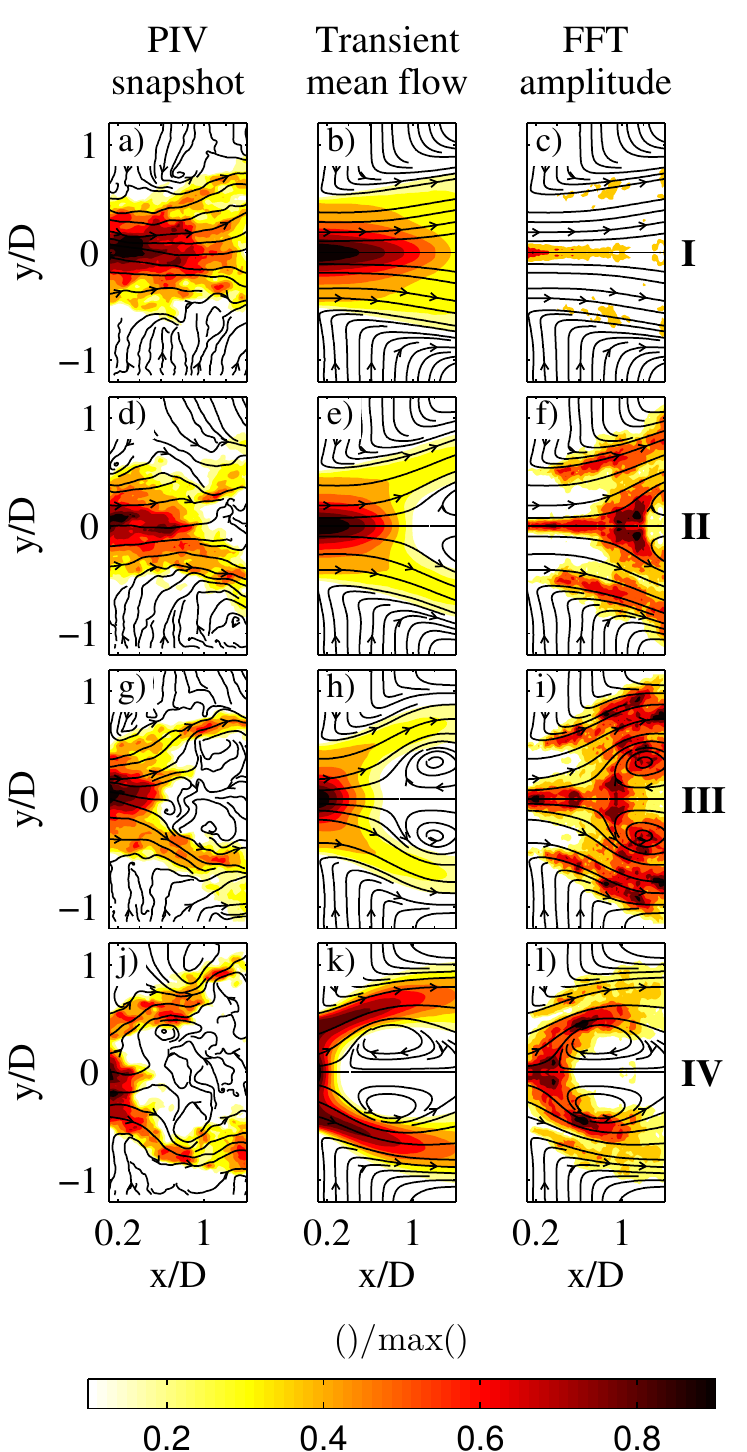}
\caption{Instantaneous axial velocity from PIV snapshots (first column), transient mean flow data (second column) and global mode amplitude form short-time FFT (right column), for four different time instances at 1.2, 1.66, 2.2 and 2.8 s (instances I-IV in \cref{fig:swirlNumber}) c).}
\label{fig:veloTotal} 
\end{figure}

We turn now to the instantaneous dynamics of the flow field during the transient. The first column of \cref{fig:veloTotal} shows the instantaneous PIV snapshots at four different instants in time, while the second column shows the corresponding base flow as reconstructed from the POD. 
The third column shows the amplitude distribution of the global mode oscillations as extracted from the short-time FFT. It indicates the flow regions where the oscillations arise during the transient. 
The four rows in \cref{fig:veloTotal} represent the different stages of the transient. Roman numerals indicate the time instances that are marked in \cref{fig:swirlNumber} c). The first row corresponds to sub-critical conditions at constant swirl, the second row to the beginning of the swirl number increase, the third row to the end of the swirl number increase, and the fourth row to the limit cycle.

It is evident from \cref{fig:veloTotal} a) and b) that the initial swirl velocity forms the characteristic overshoot in axial velocity along the center line \cite{Batchelor.1999}, but it is not sufficiently large to trigger vortex breakdown inside the measurement domain. After 1.6 s (\cref{fig:veloTotal} d and e), vortex breakdown \kilian{enters} the measurement domain from the \kilian{downstream} end. Towards the end of the swirl number increase (after 2.2 s) the vortex breakdown bubble occupies a substantial amount of the measurement domain. It moves further upstream towards the nozzle, rendering the incoming swirling jet almost annular. 

The results from the short-time FFT depicted in the third column of \cref{fig:veloTotal} reveal where and when the oscillations of the global mode arise. At the onset of the bifurcation, (\cref{fig:veloTotal} f), dominant Fourier amplitudes are detected at the upstream end of the recirculation bubble and in the incoming vortex core. This indicates the undulations of the vortex core, which is often referred to as the precession of the vortex core \cite{Syred.2006,Oberleithner.2011b}. At a later stage of the bifurcation (\cref{fig:veloTotal} i), traces of the global mode can be found in a substantial portion of the measurement domain. The oscillations in the inner and outer shear layers can be considered as the hallmark of the global mode \cite{Stohr.2012,Cala.2006}. From 2.2 s onwards, only little changes in the structure of the global mode are observed, indicating that a stable limit cycle was reached during this time period.

\moritz{\subsection{The transition scenario and the corresponding time scales}}

\begin{figure}
\centering
\includegraphics{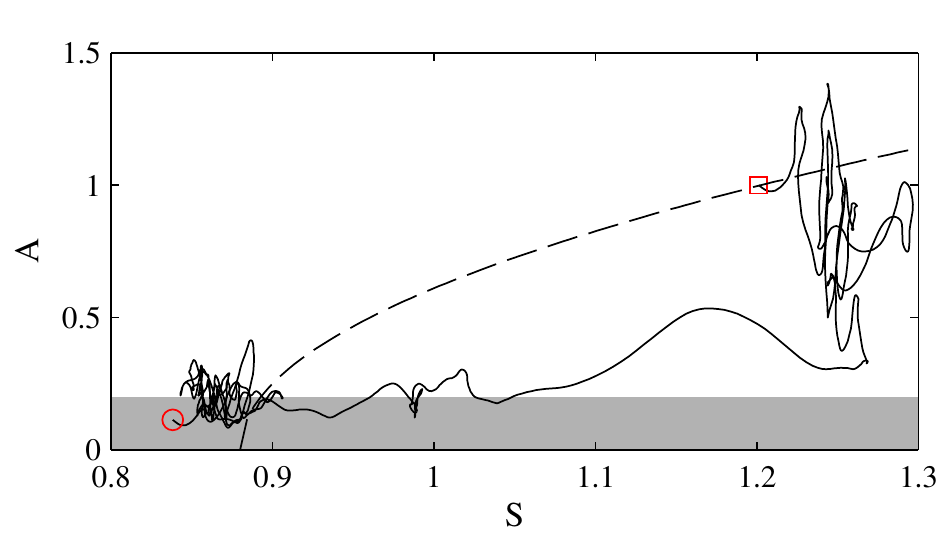}
\caption{Amplitude of the global mode versus the swirl number during the transient shown by the solid black line starting at the red circle and ending at the red square. The gray area indicates the POD noise floor and the dashed line shows the theoretical parabola of the limit cycle.}
\label{fig:A_vs_S} 
\end{figure}

\moritz{
In the introductory chapter, we proposed two different transition scenarios one named `following the limit cycle' and the other `rapid overshoot' (\cref{fig:bifDiag}). To clarify which type we encounter in the presented experiment the evolution of the amplitude of the global mode with respect to the swirl number is drawn in \cref{fig:A_vs_S}. The theoretical parabola of the limit cycle is also drawn in this figure to allow for a classification of the measured transient. \kilian{It corresponds to the limit cycle characteristic $A^* \propto \sqrt{S-S_c}$ with the critical swirl number $S_c = 0.88$, which was determined in a previous investigation\cite{Oberleithner.2011b}}. The actual mode amplitude is calculated from the corresponding two POD coefficients that represent the global mode $A = \sqrt{a_1^2 + a_2^2}$. From \cref{fig:A_vs_S} it is perfectly visible that the measured transient does not follow the limit cycle. Hence, it is almost certain that we observe a `rapid overshoot'  scenario.
}

\moritz{
Considering the stability analysis of the transient flow, it is important to discuss the different time scales of the observed phenomena. In \cref{sec:spatio-tempo} is was stated that the time scale at which the mean flow changes is directly linked to the amplitude of the global mode. 
The temporal characteristics shown in \cref{fig:swirlNumber} confirm that this is the case for the measured data. 
The graph of the swirl number, as well as the stagnation point represent the changes of the mean flow. 
They show temporal dynamics on the same timescale as the amplitude (envelope) of the global mode (order of 0.1 sec). 
The period time of the global mode (0.02 sec) lies above this time scale but both scales are not largely separated.
It is visible that the amplitude of the global mode may double during one period. 
A similar relation between the mean flow and coherent mode time scale is observed in the investigation of  Manti\u{c}-Lugo et al. \cite{ManticLugo.2015}.
There, it is shown that a mean flow stability analysis gives accurate results for this kind of transient flow.
}\\


\subsection{The temporal evolution of the region of absolute instability}

\begin{figure}
\centering
\includegraphics{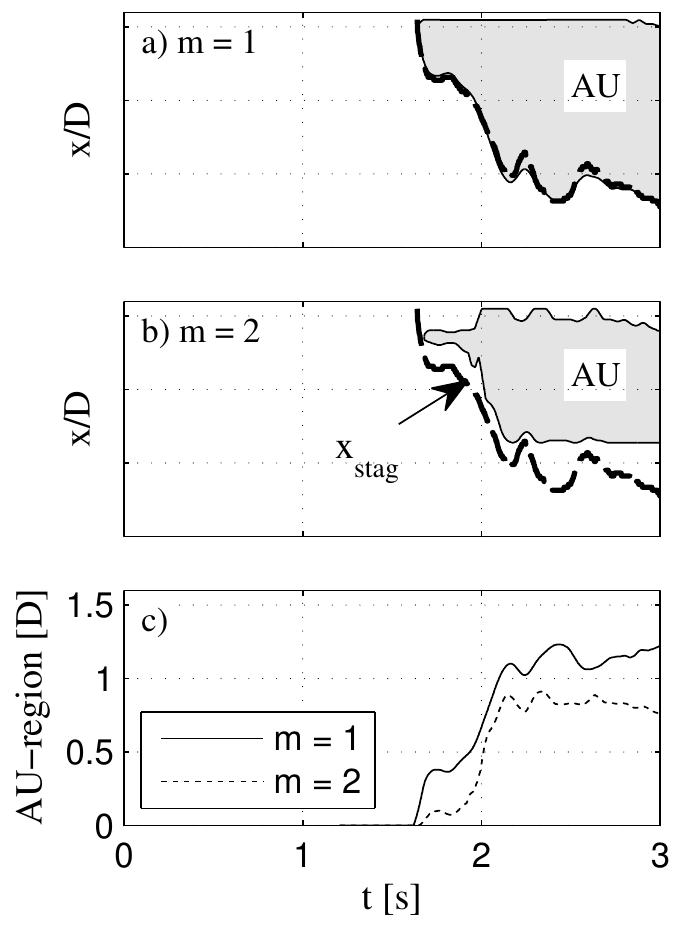}
\caption{The location of the region of absolute instability vs. time is indicated by the gray patch for mode $m = 1$ in a) and for mode $m = 2$ in b). The streamwise position of the stagnation point $x_{stag}$ is also given. The evolution of the size of the region of absolute instability for mode $m = 1$ and $m = 2$ is shown in c).}
\label{fig:stempLine} 
\end{figure}

The spatio-temporal analysis of the entire transient was carried out for the modes with azimuthal wave number $m = 1$ and $m = 2$. The analysis for $m = 2$ targets  previous experimental observations of the $m = 2$ mode at pre-breakdown conditions \cite{Liang.2005}. \Cref{fig:stempLine} a) and b) show the spatial location of the region of absolute instability for the two modes considered, while \cref{fig:stempLine} c) compares the corresponding spatial extent of the region. For the sake of comparison, the spatial location of the upstream stagnation point $x_{\mathrm{stag}}$ is included in \cref{fig:stempLine} a) and b). 

The flow becomes absolutely unstable to both modes at 1.6 s, which corresponds to the first formation of the stagnation point at 1.64 s. Note that the region of absolute instability is not confined within the measurement domain at this time step. Possibly, the flow outside the measurement domain is already absolutely unstable and the domain of absolute instability progresses with the stagnation point into the measurement domain. The extent of the absolutely unstable region steeply increases up to a time of 2.2 s and stays relatively constant in size after this time. 

The absolutely unstable domain of mode $m = 2$ shows a behavior similar to mode $m = 1$ (\cref{fig:stempLine} a) and b)). The evolution of the region of absolute instability also traces the formation and movement of the upstream stagnation point to a large extent. However,  during the initial transition phase from 1.6 to 1.8 s, the pocket of absolute instability for the mode $m = 1$ grows a lot faster than for the mode $m = 2$ (\cref{fig:stempLine}c)). At around 1.8 to 2.2 s, the region of absolute instability for mode $m = 2$ grows substantially over a very short time period. From this point in time onward, the region of absolute instability remains nearly constant in size and only changes to a minor degree with the fluctuations of the flow field.

\FloatBarrier
\subsection{Frequency selection}
\label{sec:fs}
\begin{figure}
\centering
\includegraphics{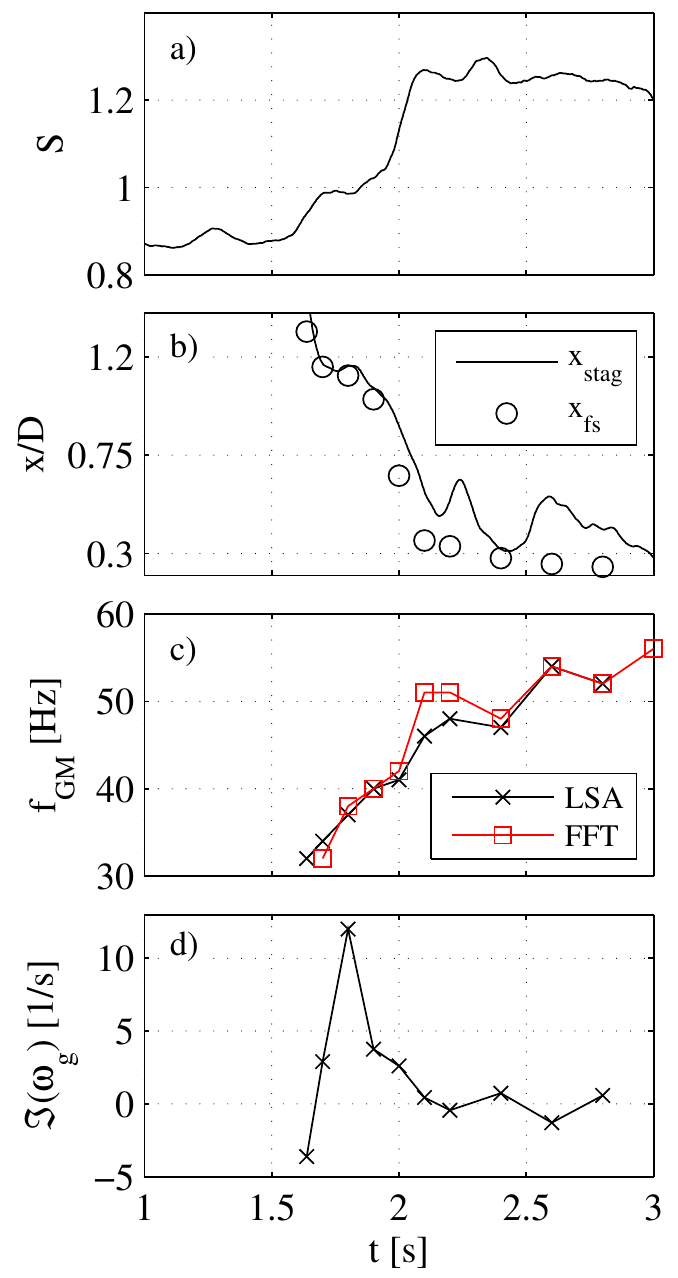}
\caption{a) Swirl number vs. time. b) The location of the wavemaker $\Re(X_{s})$ and the stagnation point $x_{stag}$ vs. time. c) Measured (FFT) and calculated (LSA) frequencies of the global mode vs. time. d) LSA prediction of the growth rate of the global mode vs. time. The roman numerals indicate representative time instances (\cref{fig:swirlNumber} c)).}
\label{fig:growthRate} 
\end{figure}

With the presence of a region of absolute instability, the flow is potentially globally unstable. 
Following step 6 of the LSA procedure, we apply the linear frequency selection criterion \cite{Chomaz.1991} to the $\omega_{0}$ curve of the $m=1$ mode.
We work out the streamwise location of the wavemaker, and determine the global mode growth rate and frequency. The results are depicted in \cref{fig:growthRate}. To put the observations in context, \cref{fig:growthRate} a) shows the evolution of the swirl number. 
The position of the wavemaker is shown in \cref{fig:growthRate} b) together with the position of the stagnation point. 
For all time steps considered, we find that the frequency selection takes place shortly upstream of the stagnation point. This agrees well with the spectral peaks of the short-time FFT shown in \cref{fig:veloTotal}. 
In \cref{fig:growthRate} c), the global mode frequency predicted by the linear frequency selection criterion is compared to the frequency derived from the measurements. The agreement is particularly good up to 2 s and from 2.4 s onward. In the range between these time instances we observe the largest deviation of predicted and computed frequencies, which is still below 10\%. Note that the first data point of the LSA calculations has no experimental counterpart, because the mode was to weak to be separated from turbulent noise at this point. The last point in the curve of experimentally obtained frequencies has no counterpart on the LSA side. The recirculation bubble had moved so close to the nozzle that the saddle point in the complex $X$-plane was pushed to negative values of $X$, i.e. inside the nozzle. Attempts to compute the analytical continuation outside the data range failed and no frequency could be obtained. 

The values of the predicted global mode growth rate are indicated in \cref{fig:growthRate} d). At 1.64 s, the first axial stations became absolutely unstable. At this point, the growth rate of the global mode is still negative. From its negative value at 1.64 s, the growth rate increases up to 1.8 s, where the largest growth rate is attained. From 1.8 to 2.2 s, the growth rate declines to zero. The growth rate predicted by LSA for times larger than 2.2 s is approximately zero, indicating that the global mode has saturated at its limit cycle.

Deriving the growth rate from the experimental data during the time from 1.6 to 2.2 s is not unambiguous due to different velocity measures. The problem of finding one POD structure that covers the entire transient evolution of the global mode has been mentioned. Taking the amplitude evolution of this structure as an indicator (\cref{fig:swirlNumber} c)) and bearing in mind that the amplitude growth exhibits a lot of jitter, a growth rate between 5 and 25 [1/s] can be derived. The largest growth rate predicted by LSA has a value of 12 [1/s] at a time of 1.8 s. This puts the LSA prediction and experimental observation in the same range, indicating that the LSA prediction is plausible. However, no conclusions can be drawn regarding the precision of the predicted growth rate.  

\FloatBarrier
\section{Discussion}
\label{discuss}
\moritz{We return now to the questions we formulated in the beginning. 
\textit{Can we apply LSA to a transient swirling flow?}
%
\kilian{Previous investigations pushed the application of LSA towards its limits and beyond and showed that it still gives accurate results.} In the present investigation we build on the confidence that the LSA was successfully applied to mean flows of turbulent swirling jets \cite{Oberleithner.2011,Oberleithner.2015c} and to transient cylinder wake flows \cite{ManticLugo.2015,Thiria.2015}. The accurate prediction of the global modes frequency throughout the entire transient shows that the principal mechanisms that drive the formation of the global mode are correctly captured. Therefore, we rely on the accurate prediction of the wave maker region.
}

\moritz{
\textit{\kilian{Where} does the global m = 1 instability arise?}
Both modes, $m = 1$ and $m = 2$, become absolutely unstable in the measurement domain at the same time. From this initial point onwards, mode $m = 1$ always shows a larger region of absolute instability (\cref{fig:stempLine} c)). Both modes are related to the formation of the internal stagnation point. The onset and strength of the absolute instability is driven by the backflow intensity. 
The frequency selection shows that the wave maker of the global mode is always located slightly upstream of the recirculation bubble. This indicates that flow conditions at the upstream stagnation point are of major importance for the formation of the global mode. The previously mentioned  emergence of a nonlinear global mode from the wake of the recalculation bubble \cite{Qadri.2013} was not found in the investigated swirling jet configuration.
}

\moritz{
\textit{\kilian{What transition scenario applies to the transient experiment?}}
As pointed out above, the data does not allow for a quantitative assessment of the growth rate predictions of the LSA. Nevertheless, a qualitative discussion of these predictions is possible and the plausibility of the LSA predictions is addressed in the following. From the different scenarios for the amplitude growth of the global mode that are conceivable during a control parameter transient that were presented in \cref{fig:bifDiag}, the experimental data suggested the `rapid overshoot' scenario. A precise quantification of the different time scales can not be obtained from the present data. However, the inspection of the experimental data indicates that the temporal changes of the mean flow and the amplitude of the global mode occur on the same time scale. 
Additional evidence for the overshoot scenario can be drawn from the amplification rate obtained from the LSA. 
The succession of growth rates shown in \cref{fig:growthRate} d), especially the positive growth rates at the early stage of the transient, are consistent with the rapid overshoot scenario.   
}



\FloatBarrier
\section{Conclusion}
\label{concl}
This study investigated the time-dependent evolution of the \kilian{global $m=1$ mode in a}  swirling jet undergoing vortex breakdown. \kilian{Local linear stability analysis was applied to the velocity field of a temporally evolving vortex breakdown bubble recorded with high-speed PIV}. Building on the experimental study of swirling flow transients by Liang \& Maxworthy \cite{Liang.2005}, it was confirmed that the onset of the global mode with azimuthal wave number one is related to the formation of a region of absolute instability. It was further confirmed that the wavemaker of the global mode is at any instant in time located upstream of the flow stagnation point. 
By comparing the global mode frequency predictions of local linear stability analysis and by relating the growth rate predictions to theoretical considerations, it was demonstrated that local linear stability analysis is capable of correctly tracing the evolution of the global mode throughout the swirl transient. Other than researchers that based their analysis on a steady but unstable solution of the Navier-Stokes equation, we find \textcolor{black}{that the global mode is not adequately described by a frequency selection criterion based on nonlinear wave front theory.}

\section*{Acknowledgements}
The financial support of the German Science Foundation (DFG) under project PA 920/29-1 and PA 920/30-1 is acknowledged. We thank Prof. V\'{a}clav Uruba for the fruitful collaboration and the loan of the high speed PIV system.
\bibliographystyle{elsarticle-num} 
\bibliography{swirlTrans}

\end{document}